\documentclass[epjST]{svjour}

\usepackage{graphicx}
\usepackage{amsmath}
\usepackage{amssymb}
\usepackage{gensymb}
\usepackage{siunitx}

\renewcommand{\epsilon}{\varepsilon}

\begin{document}

\title{Electric double layer of anisotropic dielectric colloids under
  electric fields}
\author{Ming Han,\inst{1} Huanxin Wu,\inst{2}
  \and
  Erik Luijten\inst{2,3,4}\fnmsep\thanks{\email{luijten@northwestern.edu}}}

\institute{Graduate Program in Applied Physics,
 Northwestern University, Evanston, Illinois 60208, U.S.A. 
\and Department of Physics and Astronomy, Northwestern
  University, Evanston, Illinois 60208, U.S.A.
\and Department of Materials Science and Engineering,
 Northwestern University, Evanston, Illinois 60208, U.S.A.
\and Department of Engineering Sciences and Applied Mathematics,
  Northwestern University, Evanston, Illinois 60208, U.S.A.}

\abstract{Anisotropic colloidal particles constitute an important class
  of building blocks for self-assembly directed by electrical
  fields. The aggregation of these building blocks is driven by induced
  dipole moments, which arise from an interplay between dielectric
  effects and the electric double layer. For particles that are
  anisotropic in shape, charge distribution, and dielectric properties,
  calculation of the electric double layer requires coupling of the
  ionic dynamics to a Poisson solver. We apply recently proposed methods
  to solve this problem for experimentally employed colloids in static
  and time-dependent electric fields. This allows us to predict the
  effects of field strength and frequency on the colloidal properties.}

\maketitle

\section{Introduction}

In nature, self-assembly permits simple units, from atoms to organic
molecules, to form complex structures with hierarchical order, including
various crystals and biological patterns like microtubules and
membranes.  Hierarchical organization is often the key characteristic of
functional materials~\cite{lakes93,sanchez05,aizenberg05}, determining
their properties and performance.  As a promising approach to implement
structural hierarchies in man-made materials, colloidal assembly has
received considerable
attention~\cite{velev09,fli11,vogel12,damasceno12}.  However, it is
still a formidable challenge to achieve complex structures with targeted
properties.  Isotropic spherical colloids typically assemble into
standard three-dimensional (3D) packings, i.e., hexagonal close-packed
(hcp) and face-centered cubic (fcc) lattices~\cite{pusey89}, whereas
more complex structures usually require anisotropic building blocks or
directional interactions~\cite{zhang15a}.  Synthetic colloids with
functionalized surfaces, so-called ``patchy colloids,'' are promising
candidates~\cite{zzhang04,zzhang05,lin05,hong06,glotzer07,hong08,chen11a,chen11c,ywang12,kraft12}.
Considerable efforts have been devoted to the synthesis of patchy
colloids, with a variety of approaches invented, such as glancing-angle
deposition~\cite{pawar08,chen11b}, lock-and-key
recognition~\cite{sacanna10}, cluster encapsulation~\cite{manoharan03},
and wax-droplet formation~\cite{jiang10}.  Recently, the notion of
``multivalent colloids,'' proposed in Ref.~\cite{nelson02}, was realized
experimentally~\cite{ywang12}.  Here, the colloid is a cluster of
microspheres that is self-assembled first and then ``glued'' together
via partial encapsulation, with symmetries analogous to atomic orbitals,
including $sp$, $sp^2$, $sp^3$, $sp^3d$, $sp^3d^2$, and~$sp^3d^3$.  By
programming those microspheres, e.g., by functionalizing them with
single-stranded DNA, it is possible to make them act as ``colloidal
valences.''  Like atoms, these patchy colloids can assemble into
``colloidal molecules,'' and further give access to a rich variety of
exotic materials.

In a separate development, \emph{field}-directed assembly has also been
widely reported
\cite{yethiraj03,osterman09,smallenburg12,bliu13,crassous14}.  As an
alternative approach to introduce anisotropy, external electric (or
magnetic) fields generate induced dipoles, resulting in directional
dipolar and even quadrupolar interactions between
colloids~\cite{bharti15}.  Thus, even simple spheres can form
dimension-reduced patterns in dilute dispersion, including chains along
an uniaxial alternating field~\cite{yethiraj03}, flat sheets parallel to
a biaxial rotating field~\cite{leunissen09}, and foams in a field with
magic-angle spinning~\cite{martin03}.  At high packing density, they can
form nontrivial lattices~\cite{yethiraj03} such as body-centred
tetragonal (bct) and space-filling tetragonal~(sft).  Unlike intrinsic
interactions, induced interactions require continuous energy input, but
in exchange they are much easier to adjust by simply tuning the field.
This makes it possible to create reversible and
reconfigurable~\cite{yethiraj03,crassous14} and even
self-healing~\cite{osterman09} structures.

Field-directed assembly was recently applied to patchy colloids as
well~\cite{gangwal10,yan12,yan15,song15}.  From microtubes of Janus
particles~\cite{yan12} to double helices of colloids with valences
\cite{song15}, a broad range of biomimetic patterns was discovered.
Field-directed assembly of patchy colloids exploits their anisotropic
polarizability, and thus avoids the difficulty of complicated surface
functionalization.  To rationalize and fully master the assembly, a
complete understanding of patchy colloids in external fields is
required.  The polarization of a patchy colloid is difficult to measure
in experiment due to its small size, and also analytically unsolvable
given its complex pattern.  For a colloid in a uniform dielectric
medium, the polarization can be calculated by the finite-element method
(FEM)\@.  However, for assembly directed by electric fields, colloids
are typically embedded in ionic solution, where the ions strongly
respond to the applied field and form an electric double layer (EDL)
around the colloids.  When an alternating-current (AC) field is applied,
this time-dependent spatial inhomogeneity should also be taken into
account. It is highly impractical to address this via FEM calculations
that incorporate ionic excluded-volume effects and thermal fluctuations.
The induced charges depend on the ion configuration and the external
field, and the induced charges and the external field in turn affect the
ionic distribution.

In this paper, we address this situation by systematically exploring the
response of a dielectrically anisotropic colloid and its EDL in an
aqueous electrolyte, with \emph{explicit} inclusion of salt ions and
counterions. This is made possible by a recently developed methodology
that couples molecular dynamics simulations to a dielectric solver. We
examine both static and time-dependent electric fields, and choose our
particles by analogy to those used in experiments.  We use our findings
to make experimentally important predictions on the role of field
strength and frequency and on the preferred orientations of particles
with different charge distributions and shapes.

\section{Methods and model}

\subsection{Boundary-element method}
\label{sec:method}

We perform molecular dynamics (MD) simulations of the ions surrounding
colloids that have various anisotropic shapes and charge
distributions. In each time step, the polarization of the colloid is
calculated based on the boundary-integral form of Poisson's equation,
which can be solved numerically via a boundary-element method
(BEM)~\cite{levitt78,zauhar85,allen01,boda04,jadhao12,barros14a}.  Our
implementation follows the algorithm of Ref.~\cite{barros14a}, which
gives high efficiency for MD simulations by employing a well-conditioned
formulation and solving it via a combination of the generalized minimal
residual (GMRES) method and a fast Ewald solver~\cite{barros14b}.  Here,
we extend the algorithm to incorporate an external electric field.  In
field-directed assembly experiments, the applied electric field is
typically of relative low frequency $f < 1$ MHz~\cite{gangwal10,song15}.
The permittivities of the media employed (e.g., polystyrene and water)
exhibit little variation in this frequency range~\cite{crc2015}, so that
we can assume a constant dielectric response.  Moreover, the field
applied in experiments is typically $\mathcal{O}(10^{4})$~V/m, i.e., in
the linear-response regime~\cite{butcher91}.  Consider a dielectric
interface~$\mathbf{S}$, outward unit
normal~$\hat{\mathbf{n}}(\mathbf{s})$, that separates regions of
relative permittivities $\epsilon_{\mathrm{in}}(\mathbf{s})$
and~$\epsilon_{\mathrm{out}}(\mathbf{s})$, respectively. If the surface
carries a free charge density~$\sigma_\mathrm{f}(\mathbf{s})$, the
induced charge~$\sigma_\mathrm{pol}(\mathbf{s})$ at an arbitrary surface
location can be determined by the local electric
field~$\mathbf{E}(\mathbf{s})$,
\begin{equation}
  \bar\epsilon(\mathbf{s})\left[\sigma_\mathrm{f}(\mathbf{s})
    +\sigma_\mathrm{pol}(\mathbf{s})\right]
  +\epsilon_0\mathrm{\Delta}\epsilon(\mathbf{s})\hat{\mathbf{n}}(\mathbf{s})
  \cdot \mathbf{E}(\mathbf{s})
  =\sigma_\mathrm{f}(\mathbf{s}) \;,
\label{eq:boundary}
\end{equation}
where $\epsilon_0$ is the vacuum permittivity,
$\bar\epsilon(\mathbf{s}) = [\epsilon_{\mathrm{in}}(\mathbf{s}) +
\epsilon_{\mathrm{out}}(\mathbf{s})]/2$,
and
$\mathrm{\Delta}\epsilon(\mathbf{s}) =
\epsilon_{\mathrm{out}}(\mathbf{s}) -
\epsilon_{\mathrm{in}}(\mathbf{s})$.
The electric field~$\mathbf{E}(\mathbf{s})$ consists of contributions
from both the external field~$\mathbf{E}_0(\mathbf{s})$ and all charges,
\begin{equation}
  \begin{split}
    \mathbf{E}(\mathbf{s})= \mathbf{E}_0(\mathbf{r}) &+\lim_{\delta\to
      0}\iint\displaylimits_{\mathbf{S},|\mathbf{s}-\mathbf{s}'|>\delta}
    \frac{[\sigma_{\mathrm{f}}(\mathbf{s}')+\sigma_{\mathrm{pol}}(\mathbf{s}')]
      (\mathbf{s}-\mathbf{s}')}
    {4\pi\epsilon_0|\mathbf{s}-\mathbf{s}'|^3} d\mathbf{s}'\\
    &+\iiint\displaylimits_{\mathbf{V}\setminus\mathbf{S}}
    \frac{\rho_\mathrm{f}(\mathbf{r}')(\mathbf{s}-\mathbf{r}')}
    {4\pi\epsilon_0\epsilon(\mathbf{r}')|\mathbf{s}-\mathbf{r}'|^3}
    d\mathbf{r}' \;,
    \label{eq:efield}
  \end{split}
\end{equation}
where $\mathbf{s}'$ and $\mathbf{r}'$ represent surface and off-surface
locations, respectively. $\rho_\mathrm{f}(\mathbf{r}')$ is the local
free charge density and $\epsilon(\mathbf{r}')$ represents the
corresponding permittivity.  Moreover, to avoid the divergence of the
surface integral, the infinitesimal $\delta$ is the lower bound of
$|\mathbf{s}-\mathbf{s}'|$.  Equation~\eqref{eq:boundary} must be solved
self-consistently. Within the framework of the BEM, every dielectric
interface is discretized into surface patches 
and surface charges are located at patch centers.
This transforms Eq.~\eqref{eq:boundary} into a matrix equation. Since the matrix
dimension is unchanged with the addition of an external field and the
matrix action can still be performed using an Ewald solver, we preserve
the computational efficiency of Ref.~\cite{barros14a}, which scales as
$\mathcal{O}(\mathcal{N}\log\mathcal{N})$ ($\mathcal{N}$ the total
number of free charges and surface elements) when using an Ewald solver
based on the particle--particle particle--mesh (PPPM) method.  Here,
since the system involves multiple dielectric contrasts, we utilize the
modified version of the BEM proposed in Ref.~\cite{wu16a}.  Once all
induced charges have been obtained, the forces on the ions can directly
be calculated, but the forces on the dielectric objects require
considerable care.  A detailed derivation shows that they also can be
expressed as pairwise interactions between induced and free
charges~\cite{barros14a}.

\subsection{Model}
\label{sec:model}

\begin{figure}
    \includegraphics[width=\textwidth]{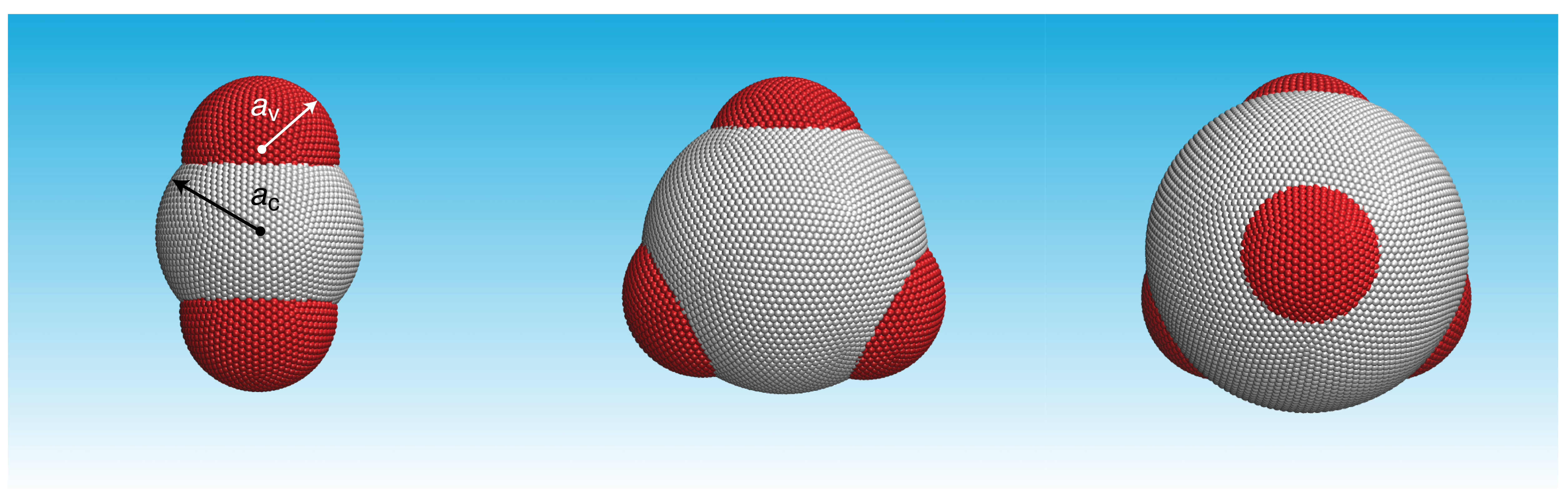}
    \caption{Two-, three-, and four-valent colloids. The red domains
      represent polystyrene ``valences'' ($n=2,3,4$) of
      radius~$a_{\mathrm{v}}$, embedded in a styrene covering (white) of
      radius~$a_{\mathrm{c}}$. The valence domains carry a negative
      electric charge whereas the covering is neutral. When suspended in
      an aqueous electrolyte, time-dependent polarization charge will be
      induced on all interfaces.  The grid points making up the surface
      are the locations at which this induced charge is computed via the
      boundary-element method.}
\label{fig:patchy_colloids}
\end{figure}

We use BEM-based molecular dynamics simulations to investigate the
polarization of colloids with valences under static and oscillating
electric fields.  The colloids are suspended in an aqueous electrolyte
at room temperature, so that the Bjerrum length characterizing the
strength of the electrostatic interactions equals
$\lambda_B = e^2/(4\pi\epsilon_0\epsilon_\mathrm{m}k_{\mathrm{B}}T) =
7.14$~\AA.
This coincides with the diameter~$\sigma$ of a hydrated ion, so that we
choose this as our length unit.  The ionic mass is the unit of mass and
all energies are expressed in units of the thermal
fluctuation~$k_{\mathrm{B}}T$, yielding the unit of time
$\tau=(m{\sigma}^2/k_{\mathrm{B}}T)^{1/2}$.  Following
Refs.~\cite{ywang12,song15}, we consider colloids with $n$ valences
($n = 2,3,4$) and a spherical covering that partially encloses the
valence spheres (Fig.~\ref{fig:patchy_colloids}).  In most of our
simulations, each valence sphere has a radius $a_\mathrm{v} = 30\sigma$.
The radius of the central sphere is chosen in accordance with the
experimental geometry~\cite{song15}, with
$a_\mathrm{c}/a_\mathrm{v} = 1.33, 1.7, 1.8$ for $n = 2, 3, 4$,
respectively.

The estimated surface charge density of the exposed surfaces of the
valence spheres is $-10$~$\mathrm{\mu}\mathrm{C}/{\mathrm{cm}}^2$ to
$-1$~$\mathrm{\mu}\mathrm{C}/{\mathrm{cm}}^2$ in aqueous
electrolyte. Accordingly, we choose a surface charge density
$\sigma_\mathrm{f}=-0.1e/{\sigma}^2$.  Surrounded by ions, the
polarization of the colloid depends not only on the dielectric response
of the colloid, but also on the structure of its EDL.  The thickness of
the EDL is typically characterized by the Debye length
$\lambda_{\mathrm{D}}=(\epsilon_0\epsilon_\mathrm{m}k_{\mathrm{B}}T/2ce^2)^{1/2}$,
where $c$ is the ionic strength of a monovalent salt solution.  In
experiment~\cite{song15}, the ionic strength $c$ ranges from $0.1$~mM to
$0.01$~mM, resulting in
$30~\mathrm{nm} \leqslant \lambda_{\mathrm{D}} \leqslant 96$~nm. Thus,
the EDL is very thin compared to the micron-sized colloids. To emulate
this size contrast, we set the Debye length to
$\lambda_{\mathrm{D}}=a_\mathrm{v}/3=10\sigma=7.14$~nm, an order of
magnitude smaller than our colloidal diameter
$d\sim4a_\mathrm{v}=120\sigma=85.6$~nm.  This requires an ionic strength
$c = 4\times10^{-4}{\sigma}^{-3}=1.8$~mM.

The BEM calculation of induced charge in this system involves three
different dielectrics: water ($\epsilon_{\mathrm{m}}=78.5$), the
polystyrene valence spheres ($\epsilon_{\mathrm{v}}=2.6$), and the
styrene covering ($\epsilon_{\mathrm{c}}=2.8$).  We discretize all
dielectric interfaces into patches of similar sizes, following the
distribution employed in Refs.~\cite{hong06,barros14b}.  Each valence
sphere is divided into 1472 patches.  For $n = 2, 3, 4$, the covering
carries $1904$, $3958$, and $4953$ patches, and encloses $632$, $970$,
and $1132$ patches of each valence sphere, respectively. Thus, 57\%,
34\%, and 23\% of each valence sphere is exposed, respectively. Although
the dielectric mismatch at the polystyrene/styrene interface is small,
we explicitly incorporate these internal surfaces when solving the
Poisson equation.  In the evaluation of the electric field at each
surface location~$\mathbf{s}$ via Eq.~\eqref{eq:efield}, the divergence
for $|\mathbf{s}-\mathbf{s}'| \to 0$ in the surface integral is
eliminated by omitting the electric field generated by the charge
located at~$\mathbf{s}$.

In the MD simulations, we examine a single colloid at a time, fixed at
the center of a cubic domain of size $300\times300\times300\sigma^3$.
Periodic boundary conditions are applied in all three dimensions.  Given
the small screening length $\lambda_{\mathrm{D}}=10\sigma$, this
simulation cell is large enough to suppress interactions between the
colloid and its periodic images.  The system contains more than
$20\,000$ ions.  All electrostatic interactions are calculated via PPPM
Ewald with a relative precision of $10^{-4}$.  In addition to
electrostatics, the ions experience excluded-volume effects as well. The
ion--ion and ion--colloid interactions are represented by expanded
shifted-truncated Lennard-Jones potentials,
\begin{equation}
  u_\mathrm{LJ}(r_{ij}) = \left\{
    \begin{array}{ll}
      \infty
      & \text{if $r_{ij}\leq \mathrm{\Delta}_{ij}$} \\
      4 k_{\mathrm{B}} T\left[(\frac{\sigma}{r_{ij}- \mathrm{\Delta}_{ij}})^{12} -
      (\frac{\sigma}{r_{ij}- \mathrm{\Delta}_{ij}})^6 +\frac{1}{4} \right]
      & \text{if $\mathrm{\Delta}_{ij} < r_{ij} <
        \mathrm{\Delta}_{ij}+2^{\frac{1}{6}}\sigma$} \\
      0
      & \text{if $r_{ij}\geq \mathrm{\Delta}_{ij}+2^{\frac{1}{6}}\sigma$}
    \end{array}\right. \;,
\label{eq:lj}
\end{equation}
where $d_i$ and $d_j$ are the diameters of particle $i$ and~$j$,
respectively, and $\mathrm{\Delta}_{ij}=(d_i+d_j)/2-\sigma$. For
ion~$i$, $d_i = \sigma$, whereas particle~$j$ is either another ion, a
valence sphere ($d_j = 2a_{\mathrm{v}}$) or the covering sphere
($d_j = 2a_{\mathrm{c}}$).  According to Ref.~\cite{barros14a}, the
average patch--patch separation must be chosen smaller than the minimum
ion--patch distance. In our system, the closest ion--surface distance is
around $0.5\sigma$. Given the large colloidal size, to avoid too large
numbers of patches, we place the patches $1.5\sigma$ inside the
colloidal surface, which allows a surface discretization with an average
patch--patch separation of approximately~$2\sigma$.

An external electric field, either direct current (DC) or alternating
current (AC), is applied to the system. Temperature is controlled via a
Langevin thermostat applied to the ions.  The damping parameter
$t_{\mathrm{d}}$ of the Langevin thermostat yields the ionic drag
coefficient $\xi=m/t_{d}$, controlling the ionic motions. Given the
Einstein relation for the diffusivity, $D=k_{\mathrm{B}}T/\xi$, the
corresponding damping parameter is $t_{\mathrm{d}}=mD/k_{\mathrm{B}}T$,
around $10^{-13}$~s in water at room temperature.
This is an order of magnitude smaller than our unit time.
To accelerate the simulations, we increase this to
$t_{\mathrm{d}} = \tau$.  Although this affects the rate of the ionic
response to the field, the polarization of the colloid mainly depends on
the competition between the field variation and the ionic response,
namely the comparison between the field frequency $f$ and the charging
frequency $f_{\mathrm{c}}$ of the EDL (see Sec.~\ref{sec:dipole-ac}).
All simulations start from a random initial ionic configuration and run
for $5\times 10^5$ steps (for a DC field) or 20 AC cycles ($3\times10^5$
to $3\times10^6$ steps, depending on frequency), with timestep
$dt = 0.01\tau$.  For all parameter choices, the ionic configurations
reach a representative state after $10^4$ steps, and each run yields 200
independent samples.

\section{Results}

Through detailed MD simulations coupled to the dielectric solver as
described in Sec.~\ref{sec:method}, we examine the dependence of the
polarization on the strength, frequency, and orientation of the external
field. We employ the model described in Sec.~\ref{sec:model}, and first
perform a quantitative study for a colloid with $n=2$ valences, and
then generalize this to a qualitative study of $n=3$ and~$n=4$.

\subsection{Effect of a static electric field on the electric double
  layer}

\begin{figure}
  \centerline{\includegraphics[width=0.85\textwidth]{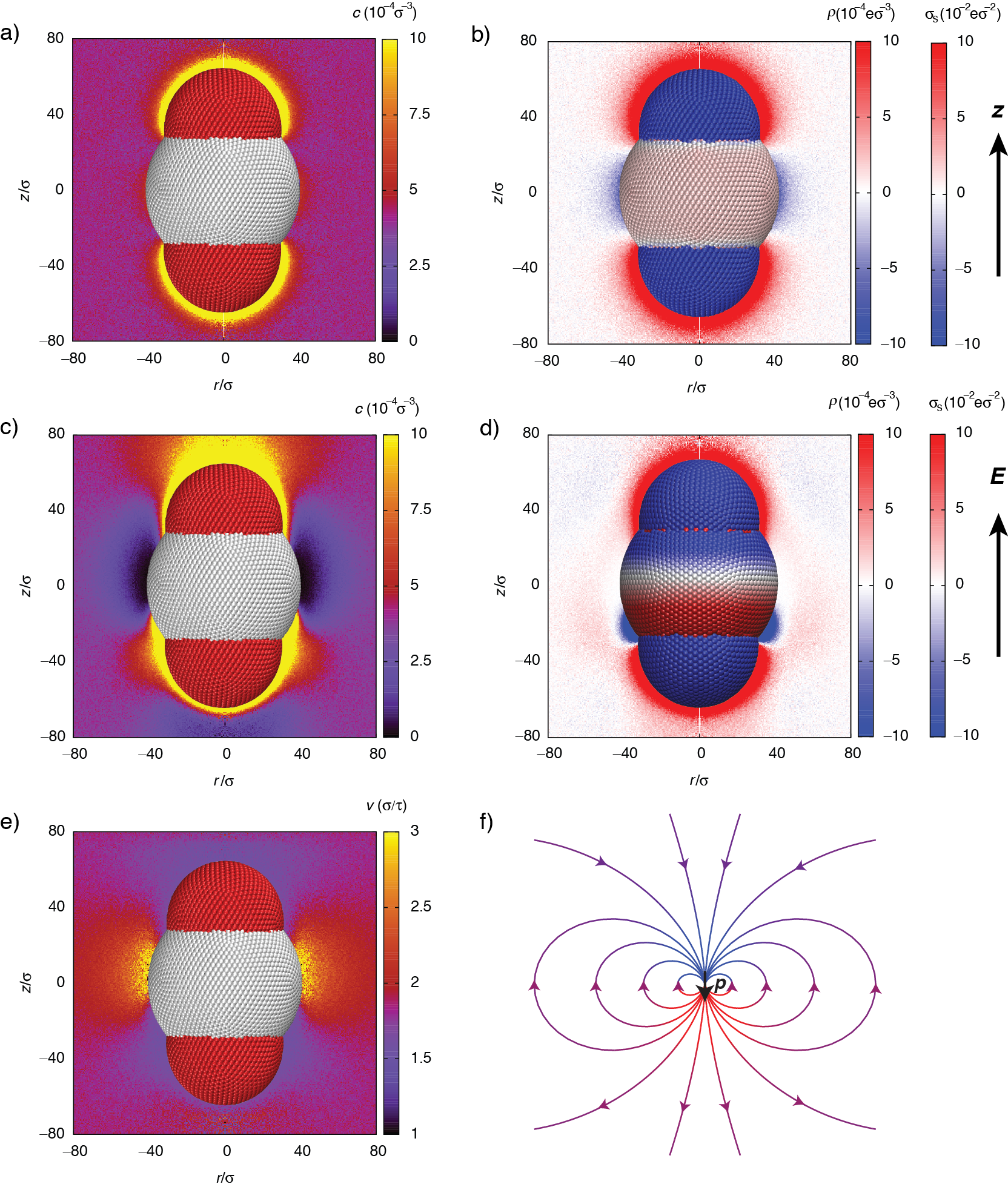}}
  \caption{Influence of a DC field on the electric double layer of a
    colloid with two charged ``valence'' domains, embedded in an aqueous
    electrolyte. Top row shows the case without an external
    field. a)~The ion concentration $c(r,z)$ (in cylindrical
    coordinates) is significantly enhanced near areas of high surface
    charge, and otherwise uniform at the bulk
    concentration~$c = 4 \times 10^{-4}\sigma^{-3}$. b)~The
    corresponding charge distribution, both due to the ion distribution
    and due to the free and induced surface charge
    density~$\sigma_\mathrm{s}$ on the colloid. The negative free
    surface charge on the exposed areas of the valence domains is
    responsible for the positive induced charge at the interface between
    the central spherical covering and the surrounding water. Middle row
    shows the corresponding panels for the same colloid in an external
    DC field, applied parallel to the long axis of the colloid and of
    strength $E = 0.2k_{\mathrm{B}}T/e\sigma$.  c)~The ion distribution
    is now strongly distorted near the valence domains, and depleted in
    the equatorial region around the central covering. As explained in
    the main text, this arises from the high ion velocities (speed
    profile in panel~e)) that in turn result from the enhancement of the
    applied electric field by the induced dipolar field of the
    particle. d)~The polarization charge on the central spherical region
    and the corresponding ionic charge distribution both reflect the
    symmetry breaking imposed by the external field, resulting in a net
    dipole moment on the particle.  f)~Electric field generated by a
    point dipole~$\mathbf{p}$ oriented oppositely to the applied
    field~$\mathbf{E}$.  The streamlines are color coded based upon the
    local electrostatic potential arising from the dipole.  For visual
    clarity, in panels b) and~d) the color scale for $\sigma_\mathrm{s}$
    saturates at $\sigma_{\mathrm{f}} = \pm0.1e/{\sigma}^2$.}
\label{fig:charge_distr}
\end{figure}

Figure~\ref{fig:charge_distr} summarizes the effects of an external
electric field on the EDL of a divalent colloid. As expected, in the
absence of an electric field, its two charged valences play a dominant
role in the EDL structure, as they strongly attract counterions
(Fig.~\ref{fig:charge_distr}a,b).  The resultant EDL has a thickness
$\lambda_{\mathrm{D}} = 10\sigma$ and an ionic concentration several
times higher than the bulk concentration
$c = 4\times10^{-4}\sigma^{-3}$. In addition, the negative charge on the
valence domains (red surface areas in Fig.~\ref{fig:charge_distr}a)
generates an electric field on the styrene covering (white surface areas
in Fig.~\ref{fig:charge_distr}a) that points inward. As the covering has
a lower permittivity ($\epsilon=2.8$) than the surrounding medium
($\epsilon=78.5$), this induces a positive charge on the originally
neutral surface. The induced charge then attracts negative ions
(Fig.~\ref{fig:charge_distr}b).

These charge distributions are strongly affected by a symmetry-breaking
direct-current (DC) field of strength
$E = 0.2 k_{\mathrm{B}}T/(e\sigma)$ applied along the long axis of the
colloid.  Generally, such a DC field will induce electrophoresis of this
net negatively charged colloid~\cite{long98}. However, here we merely
consider the case of a colloid held fixed in place, to obtain a general
picture of the effects of external electric fields on the EDL\@.  The
EDL near both valence domains is distorted in the direction of the field
(Fig.~\ref{fig:charge_distr}c).  At the same time, the styrene covering
now exhibits an induced surface charge that varies from positive to
negative in the field direction, resulting in an effective dipole moment
antiparallel to the field (Fig.~\ref{fig:charge_distr}d). Unlike the
zero-field case, the induced charge does not lead to an accumulation of
counterions around the entire covering. Instead,
Fig.~\ref{fig:charge_distr}c demonstrates a depletion of ions (dark
regions)---an observation that can be understood from the effective
dipole moment. Namely, a point dipole~$\mathbf{p}$ generates an electric
field opposite to itself in the plane perpendicular to~$\mathbf{p}$
(Fig.~\ref{fig:charge_distr}f). Thus, in this region the polarized
covering enhances the applied external field, resulting in a strong
driving force on the ions. This is confirmed in the speed profile,
Fig.~\ref{fig:charge_distr}e.  Once the system has reached a steady
state, continuity requires that the ionic current density
$\mathbf{j} = c\mathbf{v}$ integrated over any plane perpendicular to
the $z$-axis must be independent of~$z$.  Thus, higher ionic velocities
imply lower concentrations.  Moreover, ionic motion typically leads to
electroosmotic flows around a fixed, charged colloid. Here, the
polarization of the colloid clearly can have a significant impact on
electroosmosis, which only recently has been considered
theoretically~\cite{levitan05,bazant06,squires06,bazant09}.  Once the
particle is released, it will move oppositely to the DC field.  Although
this electrophoresis of a single colloid is not affected by dielectric
effects owing to the particle symmetry and the uniformity of the field,
collective dynamics of multiple colloids should be affected. Such
electrokinetic effects will be the subject of future work.

\subsection{Minimum field strength required for a stable polarization}

\begin{figure}
  \includegraphics[width=\textwidth]{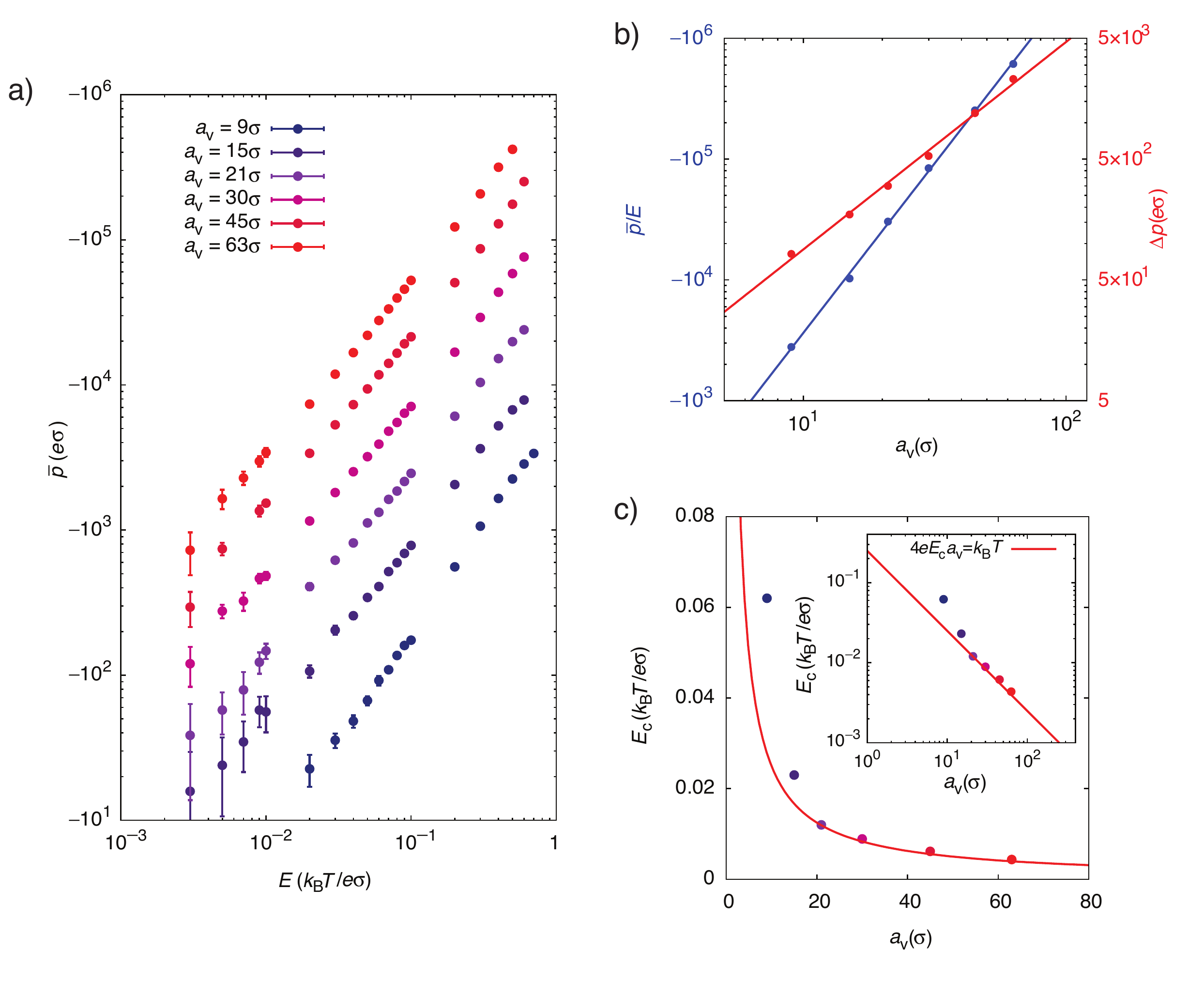}
  \caption{Effects of DC field strength on the polarization of a colloid
    with two valences (cf.\ Fig.~\ref{fig:charge_distr}) in an aqueous
    electrolyte.  a) Ensemble-averaged dipole moment~$\bar{p}$ of the
    electric double layer as a function of field strength for six
    different colloid sizes (indicated as the size~$a_{\mathrm{v}}$ of
    the valence domains, but note that the aspect ratio
    $a_{\mathrm{c}}/a_{\mathrm{v}}$ is kept constant) on a double
    logarithmic scale.  b)~Dependence of the ratio $\bar{p}/E$ (blue,
    evaluated at $E=0.2k_{\mathrm{B}}T/e\sigma$) and the standard
    deviation $\mathrm{\Delta}p$ (red) on the radius of a valence sphere
    $a_\mathrm{v}$.  Errorbars are smaller than the data points. The
    fitted curves for $\bar{p}/E$ and $\mathrm{\Delta}p$ have power-law
    exponents $2.8$ and~$1.7$, respectively.  c)~Threshold
    field~$E_\mathrm{c}$ below which the standard
    deviation~$\mathrm{\Delta}p$ exceeds $\bar{p}$, as a function of
    $a_\mathrm{v}$ (replotted logarithmically in the inset). The curve
    $4eE_\mathrm{c}a_{\mathrm{v}}=k_{\mathrm{B}}T$ is the theoretical
    prediction.}
\label{fig:threshold}
\end{figure}

In the absence of ions, the polarization of an object is based on its
dielectric response, with an induced dipole moment~$p$ that is (for
fields that are not too strong) linearly proportional to the field
strength~$E$ and the volume~$V$ of the object, $p \propto
EV$~\cite{jackson99}.  In aqueous electrolytes, the situation is more
complicated, as the polarization also depends on the ion
distribution in the EDL\@.  Experimentally, too strong electric fields
are generally disfavored, to avoid Faradaic reactions that can lead to
electrode degradation or dissolution, as well as bubble
formation~\cite{levitan05,bazant06,squires06}. On the other hand, for
weak electric fields, thermal fluctuations of the ions can overwhelm
the dielectric response of the colloid and cause large variations in
the induced dipole moment.  Thus, an important question concerns the
minimum field strength required for a stable polarization, and the
dependence of this threshold on particle size.

We investigate these questions by systematically varying the field
strength over three orders of magnitude, and by comparing colloids of
six different colloid sizes (with
$a_{\mathrm{v}}=9\sigma, 15\sigma, 21\sigma, 30\sigma, 45\sigma,
63\sigma$),
all with the same aspect ratio $a_{\mathrm{c}}/a_{\mathrm{v}}=1.33$.  As
before, the field is applied along the long axis of the colloid.  To
quantify the polarization, we calculate the net dipole moment~$p$ of the
induced surface charges on the colloid and ions inside its EDL (which we
define here as the region within $\lambda_{\mathrm{D}} = 10\sigma$ from
the colloidal surface). Note that the free surface charge does not
contribute to the dipole moment due to spatial symmetry.  As illustrated
in Fig.~\ref{fig:threshold}a, for sufficiently strong fields, the
ensemble-averaged dipole moment~$\bar{p}$ is indeed linear in the field
strength~$E$. We evaluate the ratio~$\bar{p}/E$ for these six
different-sized colloids at $E = 0.2k_{\mathrm{B}}T/e\sigma$, where all
of them exhibit clear linearity of their polarizations with the field
strength, and observe that it follows a power-law dependence on the
valence radius~$a_{\mathrm{v}}$, with an exponent almost equal to $3$
(Fig.~\ref{fig:threshold}b).  This confirms that $\bar{p} \propto EV$.
We characterize the thermal flucutations of the dipole moment by its
standard deviation $\mathrm{\Delta}p$, and find that this also displays
a power-law dependence on $a_{\mathrm{v}}$, with an exponent~$1.7$
(Fig.~\ref{fig:threshold}b).  If the smaller particles, with colloid
size close to the Debye length, are omitted, the dependence
on~$a_{\mathrm{v}}$ is quadratic.  Indeed, if we estimate that the EDL
contains $N$ ions, then owing to thermal fluctuations they may have an
imbalance between the two poles of the colloid
$\mathrm{\Delta}N \propto \sqrt{N}$.  This would result in an
instantaneous dipole moment $\sqrt{N}d$, where $d$ is the colloidal
size.  At weak field strength, the number of ions is approximately
$N = cA\lambda_{\mathrm{D}}$, where $A \propto d^2$ is the colloidal
surface area.  Thus, the resultant variations in dipole moment
$\mathrm{\Delta}p$ are proportional to
$\sqrt{N}d \propto d^2 \propto a_{\mathrm{v}}^2$.

For each colloidal size, there exists a threshold $E_{\mathrm{c}}$ in
the field strength below which the variations
$\mathrm{\Delta}p > \bar{p}$ make the dipole moment $p$ unpredictable.
Given the size dependences $\bar{p} \propto Ea^3_{\mathrm{v}}$ and
$\mathrm{\Delta}p \propto a^2_{\mathrm{v}}$, we expect this threshold
$E_{\mathrm{c}}$ to be inversely proportional to $a_{\mathrm{v}}$.
Indeed, as illustrated in Fig.~\ref{fig:threshold}c, our results satisfy
$eE_{\mathrm{c}}l = k_{\mathrm{B}}T$, where $l = 4a_{\mathrm{v}}$ is the
colloidal length along the applied field.  This implies a general rule
for a stable polarization: the potential drop along the colloid due to
the applied field should be larger than the thermal fluctuations
$k_{\mathrm{B}}T$.  Furthermore, the deviations of $E_{\mathrm{c}}$ from
this requirement for small colloidal sizes indicate that an even
stronger field is required if the Debye length is comparable to or
larger than the colloidal size, e.g., at low ionic concentrations or for
small colloids.

\subsection{Effective dipole moment in an AC field}
\label{sec:dipole-ac}

In experiment, applying a DC field can be problematic: the electrodes
are usually strongly screened by counterions and thus generate weak
fields in the bulk.  This is typically resolved by applying an AC field
instead.  For field-directed assembly~\cite{song15}, the commonly used
AC frequency range is between $100$~kHz and $1$~MHz, at which ions move
locally and barely cause global flows, so that electro-hydrodynamics can
be ignored~\cite{bazant09}.  In this frequency range, all dielectrics
respond to the field instantaneously, whereas the evolution of the EDL
due to ionic motion exhibits a strong frequency
dependence~\cite{mangelsdorf97}.

The dynamic ion exchange between the EDL and the bulk electrolyte can be
analyzed through analogy with an RC circuit~\cite{squires06}.  Under an
electric field, the EDL of a spherical colloid of radius~$a$
($a > \lambda_{\mathrm{D}}$) act as a capacitor with
$C \approx 4\pi a^2\epsilon_0\epsilon_{\mathrm{m}}/
\lambda_{\mathrm{D}}$,
whereas the bulk electrolyte behaves as a resistor with
$R \approx 1/4 \pi a\sigma_\mathrm{b}$, where $\sigma_{b}$ is the bulk
ionic conductivity. Thus, the EDL has a charging time
$\tau_{\mathrm{c}} = RC \approx
a\epsilon_0\epsilon_{\mathrm{m}}/\sigma_\mathrm{b}\lambda_{\mathrm{D}}$,
which corresponds to a characteristic frequency
$f_{\mathrm{c}} = 1/\tau_{\mathrm{c}}$.  If the AC field frequency~$f$
is much lower than $f_{\mathrm{c}}$, the EDL will have enough time to
adjust and remain in phase with the field.  However, if $f$ approaches
or exceeds $f_{\mathrm{c}}$, the structure of the EDL structure cannot
fully track the field.

\begin{figure}
  \includegraphics[width=\textwidth]{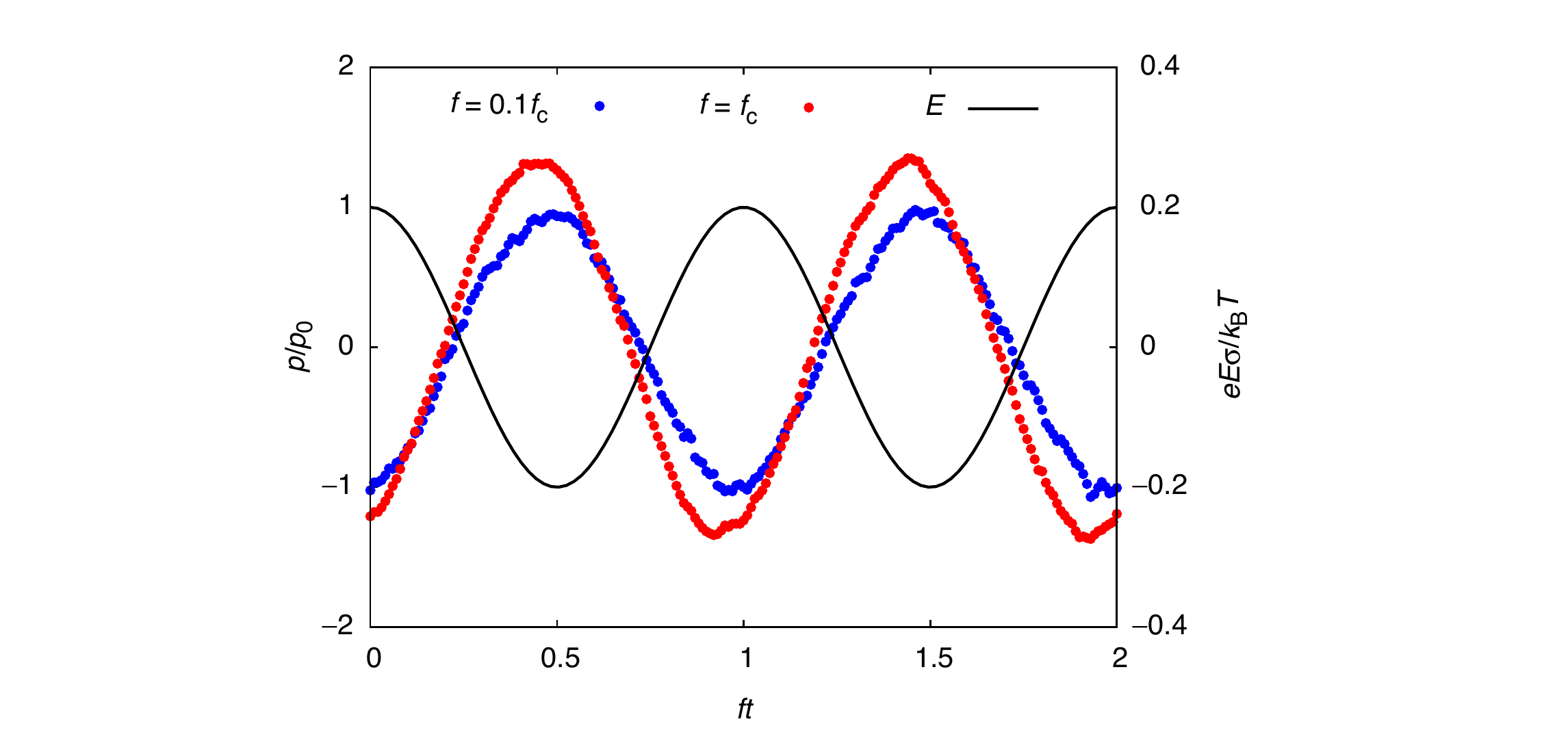}
  \caption{Time-dependent polarization of a colloid with two valences
    under a sinusoidal AC field (black solid curve) with amplitude
    $E_0 = 0.2k_{\mathrm{B}}T/e\sigma$.  At low frequency
    $f = 0.1f_{\mathrm{c}}$ (blue), with
    $f_{\mathrm{c}} = D/\lambda_{\mathrm{D}}a_{\mathrm{v}}$, the net
    dipole moment of the surface charge and the diffuse ion layer in
    perfect antiphase with the field. At $f = f_{\mathrm{c}}$ (red) the
    polarization is enhanced due to incomplete screening and exhibits a
    phase shift.  For both frequencies, the dipole moment~$p$ is scaled
    by the static ($f=0$) polarization~$p_0$.}
\label{fig:frequency}
\end{figure}

Here we apply a sinusoidal AC field $E = E_0 \cos(2\pi f t)$, with
$E_0 = 0.2 k_{\mathrm{B}}T/e\sigma$, to investigate the frequency
dependence of the dynamic polarization of a colloid with two valences.
To compare with the static polarization ($f = 0$), we scale the
time-dependent dipole moment~$p$ by the ensemble-averaged dipole moment
$p_0$ under a DC field~$E_0$.  The conductivity of a monovalent salt
solution is $\sigma_\mathrm{b} = 2ce^2/\xi$, where $\xi$ is the drag
coefficient of an ion.  Given the relation between $\xi$ and the
diffusivity~$D$ (cf.\ Sec.~\ref{sec:model}), we estimate the charging
frequency of the EDL as
$f_{\mathrm{c}} = \sigma_\mathrm{b}\lambda_{\mathrm{D}} /
a_{\mathrm{v}}\epsilon_0\epsilon_{\mathrm{m}} =
D/\lambda_{\mathrm{D}}a_{\mathrm{v}}$.
As illustrated in Fig.~\ref{fig:frequency}, at a low frequency
$f = 0.1f_{\mathrm{c}}$, the dipole moment~$p$ is in antiphase with the
field and its amplitude is close to the static polarization case~$p_0$.
However, at a higher frequency $f = f_{\mathrm{c}}$, the EDL lacks time
to fully form.  Thus, the screening effects on the colloid become weaker
and the total polarization~$p$ is enhanced.  Meanwhile, since the EDL
that compensates the polarization has a phase lag compared to the
applied field, $p$ is ahead of the field.  Returning to the experiments
of Ref.~\cite{song15}, we note that micron-sized colloids are immersed
in a solution with Debye length $\lambda_\mathrm{D} \sim 100$~nm and
exposed to an AC field of frequency $f = 100$~kHz.  Given the ionic
diffusivity in water $D \sim 5\times 10^{-9}$
${\mathrm{m}}^2/\mathrm{s}$~\cite{harned47}, the charging frequency of
the EDL is then $f_{\mathrm{c}} \sim 50$~kHz.  Thus, the polarizations
of these colloids fall in the high-frequency regime
($f \geq f_{\mathrm{c}}$).

\subsection{Orientational preference}

\begin{figure}
    \includegraphics[width=\textwidth]{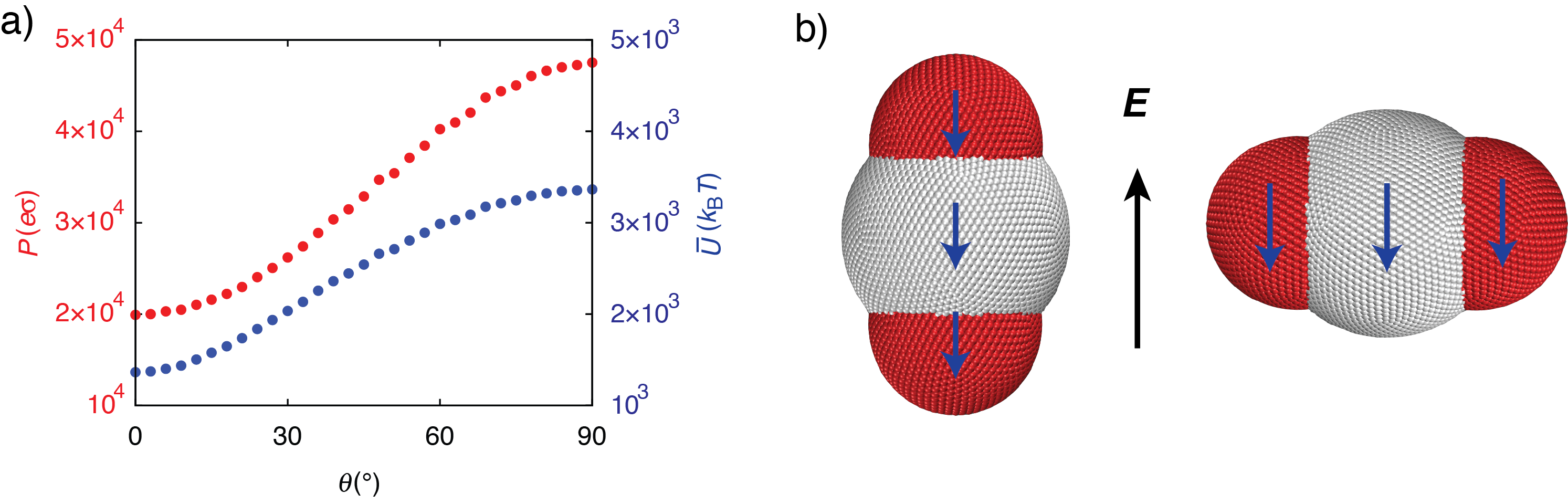}
    \caption{Orientational preference of a colloid with two valences
      under an AC field of frequency $f = f_{\mathrm{c}}$ and strength
      $E = 0.2k_{\mathrm{B}}T/e\sigma$.  a)~Amplitude of the induced
      dipole moment~$P$ (red) and the time-averaged electrostatic
      energy~$\bar{U}$ (blue, average of the instantaneous electrostatic
      energy $U=-\mathbf{p}\cdot\mathbf{E}$ over 15 AC cycles) as a
      function of the tilt angle~$\theta$ between the long axis of the
      colloid and the external field.  b)~Schematic illustration of the
      effective dipoles of the two valence domains and the central
      covering, for $\theta = 0 \degree$ and $\theta = 90 \degree$,
      respectively.}
\label{fig:p-theta}
\end{figure}

In field-directed assembly, the orientations of anisotropic building
blocks play an important role in determining the resulting
structure~\cite{zerrouki08,crassous14,song15}.  To investigate the
orientational preferences of a colloid of valence two, we fix the
colloid and systematically vary the tilt angle of the applied field with
respect to the long axis of the colloid.  In accordance with experiment,
the applied field is chosen in the high-frequency regime, with
$f = f_{\mathrm{c}}$.  We characterize the polarization of the colloid
under this AC field via the amplitude $P$ of its induced dipole moment.
As illustrated in Fig.~\ref{fig:p-theta}a, the polarization is enhanced
with the tilt angle~$\theta$. Usually, for a polarizable object with an
elongated shape the separation between its positive and negative
polarized surface charges becomes largest when the object is aligned
with the external field, resulting in the strongest polarization. By
contrast, here we find that the perpendicular configuration
($\theta=90 \degree$) yields the strongest polarization. This can be
understood from a simple dipole analysis.  The colloid contains three
dielectric domains, namely the two valences and the covering.  As
illustrated in Fig.~\ref{fig:p-theta}b, for both the parallel
($\theta = 0 \degree$) and perpendicular ($\theta = 90 \degree$)
configurations, upon polarization, the effective dipole moments of all
these domains are opposite to the applied field~$\mathbf{E}$, and thus
parallel to each other.  The electric field at position $\mathbf{r}$
exerted by a dipole $\mathbf{p}$ at the origin is
\begin{equation}
  \mathbf{E}_p =
  \frac{1}{4\pi{\epsilon}_0{\epsilon}_{\mathrm{m}}}
  \left(\frac{3\left(\mathbf{p} \cdot
        \hat{\mathbf{r}}\right)\hat{\mathbf{r}} - \mathbf{p}}{r^3}\right)
  - \frac{1}{3{\epsilon}_0{\epsilon}_{\mathrm{m}}}
  \mathbf{p}{\delta}^3(\mathbf{r})\;,
  \label{eq:dipole-field}
\end{equation}
where ${\delta}^3(\mathbf{r})$ is the 3D Dirac delta function. This
field reduces to
$\mathbf{E}_p = \mathbf{p}/(2\pi{\epsilon}_0{\epsilon}_{\mathrm{m}}r^3)$
when $\mathbf{r}$ is parallel to $\mathbf{p}$, whereas
$\mathbf{E}_p =
-\mathbf{p}/(4\pi{\epsilon}_0{\epsilon}_{\mathrm{m}}r^3)$
when $\mathbf{r}$ is perpendicular to~$\mathbf{p}$.  When the colloid is
parallel to the field ($\theta = 0 \degree$), the effective dipole on
each domain generates an electric field on each of the other domains
that counteracts the applied field, and suppresses the polarization.  By
contrast, in the perpendicular configuration of the colloid, the dipole
moments generate fields aligned with the external field and thus enhance
the polarization.  Given that in both configurations the net dipole
moment is oriented oppositely to the applied field, the parallel
configuration, with the smallest dipole moment, is energetically
preferred, as confirmed by the dependence of the time-averaged
electrostatic energy~$\bar{U}$ of the colloid on the tilt angle~$\theta$
(Fig.~\ref{fig:p-theta}a).  This also explains the experimental
observation that colloids with $n=2$ orient along the
field~\cite{song15}. Interestingly, if the colloid has a higher permittivity
than the medium, the induced dipoles on the domains are parallel to 
the applied field, and according to the previous analysis
the parallel configuration leads to the \emph{strongest}
polarization. Yet, again the parallel orientation is energetically
preferred.

\subsection{Colloids with three and four valences}

\begin{figure}
  \includegraphics[width=\textwidth]{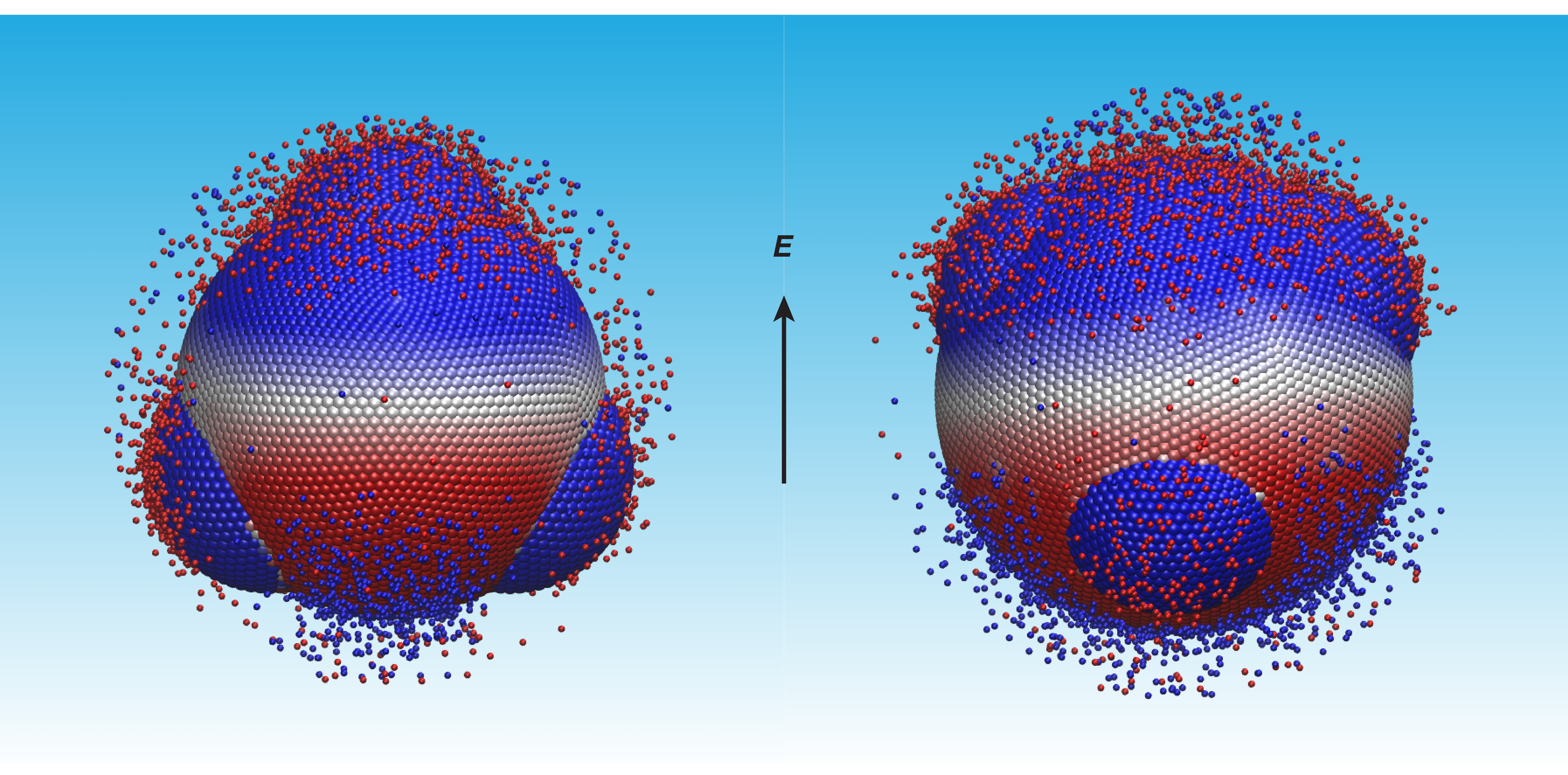}
  \caption{Polarization and preferred orientation colloids with $n=3$
    (left) and $n=4$ (right) valences, when exposed to a DC field of
    strength $E = 0.2 k_{\mathrm{B}}T/e\sigma$ in an aqueous
    electrolyte.  Snapshots of their steady states are shown, with the
    colloids color coded according to their surface charge density.  The
    surrounding ions are shown up to $2\lambda_{\mathrm{D}} = 20 \sigma$
    away from the colloidal surfaces.}
  \label{fig:three-four}
\end{figure}

The above findings on the polarization of a colloid with two valences
can be generalized to the cases of three and four valences colloids.
Here, we provide a first qualitative exploration of these particles.  We
expose the colloid to a DC field of strength
$E = 0.2 k_{\mathrm{B}}T/e\sigma$ and, unlike for $n=2$, allow the
colloid to freely rotate about its center, to determine its
energetically preferred orientation.

Figure~\ref{fig:three-four} demonstrates these orientations for colloids
with both $n=3$ and $n=4$.  The central covering, which is neutral in
the absence of a field, exhibits an induced surface charge distribution
that effectively generates a dipole opposite to the field.  However, the
surface charges of the valence domains remain dominated by their
original free charges. This figure also shows a snapshot of the
arrangement of ions near the colloid (within
$2\lambda_\mathrm{D} = 20 \sigma$ from the surface). Unsurprisingly,
counterions aggregate near and screen the charged surface areas. However,
the ion concentration near the central covering is depleted, owing to
the same mechanism as demonstrated in Fig.~\ref{fig:charge_distr}.

\section{Summary}

In summary, using a newly introduced, highly efficient algorithm that
couples molecular dynamics simulations with a dielectric solver, we have
examined the polarization and electric double layer of ``colloids with
valences.'' The calculations rely on some unique capabilities:
particle-based simulations make it possible to examine fluctuation and
excluded-volume effects of dielectric systems, and the
boundary-element-based Poisson solver makes it possible to examine
arbitrarily shaped dielectric interfaces. The particles investigated
were inspired by experimental work on field-directed assembly. In
addition to providing insight into the induced surface charge
distributions of these building blocks and the surrounding ion clouds,
we have also explicitly demonstrated that the induced dipole moment is
linear in the applied field and proportional to the particle
volume. Moreover, at weak fields the thermal fluctuations start to
dominate; we established that the threshold field strength for a stable
polarization is inversely proportional to particle size. For AC fields,
the preferred orientation is parallel to the field, despite the fact
that in this case the induced dipole moment is smallest.  Whereas these
conclusions were drawn for colloids with two valence domains, we
provided qualitative evidence that they extend to $n=3$ and $n=4$
colloids as well. This work opens pathways toward predictive
capabilities for polarizable building blocks for the field-directed
assembly of new materials.

\acknowledgement
This research was supported through award 70NANB14H012 from the
U.S. Department of Commerce, National Institute of Standards and
Technology, as part of the Center for Hierarchical Materials Design
(CHiMaD), the National Science Foundation through Grant Nos.\
DMR-1121262 at the Materials Research Center of Northwestern University
and DMR-1310211, and the Center for Computation and Theory of Soft
Materials (CCTSM) at Northwestern University.  We thank the Quest
high-performance computing facility at Northwestern University for
computational resources.


\begin{thebibliography}{56}

\bibitem{lakes93}
R.~Lakes, Nature \textbf{361}, 511 (1993)

\bibitem{sanchez05}
C.~Sanchez, H.~Arribart, M.M. Giraud~Guille, Nature Mater. \textbf{4}, 277
  (2005)

\bibitem{aizenberg05}
J.~Aizenberg, J.C. Weaver, M.S. Thanawala, V.C. Sundar, D.E. Morse, P.~Fratzl,
  Science \textbf{309}, 275 (2005)

\bibitem{velev09}
O.D. Velev, S.~Gupta, Adv. Mater. \textbf{21}, 1 (2009)

\bibitem{fli11}
F.~Li, D.P. Josephson, A.~Stein, Angew. Chem. Int. Ed. \textbf{50}, 360 (2011)

\bibitem{vogel12}
N.~Vogel, C.K. Weiss, K.~Landfester, Soft Matt. \textbf{8}, 4044 (2012)

\bibitem{damasceno12}
P.F. Damasceno, M.~Engel, S.C. Glotzer, Science \textbf{337}, 453 (2012)

\bibitem{pusey89}
P.N. Pusey, W.~van Megen, P.~Bartlett, B.J. Ackerson, J.G. Rarity, S.M.
  Underwood, Phys. Rev. Lett. \textbf{63}, 2753 (1989)

\bibitem{zhang15a}
J.~Zhang, E.~Luijten, S.~Granick, Ann. Rev. Phys. Chem. \textbf{66}, 581 (2015)

\bibitem{zzhang04}
Z.~Zhang, S.C. Glotzer, Nano Letters \textbf{4}, 1407 (2004)

\bibitem{zzhang05}
Z.~Zhang, A.S. Keys, T.~Chen, S.C. Glotzer, Langmuir \textbf{21}, 11547 (2005)

\bibitem{lin05}
S.~Lin, M.~Li, E.~Dujardin, C.~Girard, S.~Mann, Adv. Mater. \textbf{17}, 2553
  (2005)

\bibitem{hong06}
L.~Hong, A.~Cacciuto, E.~Luijten, S.~Granick, Nano Letters \textbf{6}, 2510
  (2006)

\bibitem{glotzer07}
S.C. Glotzer, M.J. Solomon, Nature Mater. \textbf{6}, 557 (2007)

\bibitem{hong08}
L.~Hong, A.~Cacciuto, E.~Luijten, S.~Granick, Langmuir \textbf{24}, 621 (2008)

\bibitem{chen11a}
Q.~Chen, J.K. Whitmer, S.~Jiang, S.C. Bae, E.~Luijten, S.~Granick, Science
  \textbf{331}, 199 (2011)

\bibitem{chen11c}
Q.~Chen, S.C. Bae, S.~Granick, Nature \textbf{469}, 381 (2011)

\bibitem{ywang12}
Y.~Wang, Y.~Wang, D.R. Breed, V.N. Manoharan, L.~Feng, A.D. Hollingsworth,
  M.~Weck, D.J. Pine, Nature \textbf{491}, 51 (2012)

\bibitem{kraft12}
D.J. Kraft, R.~Ni, F.~Smallenburg, M.~Hermes, K.~Yoon, D.A. Weitz, A.~van
  Blaaderen, J.~Groenewold, M.~Dijkstra, W.K. Kegel, Proc. Natl. Acad. Sci.
  U.S.A. \textbf{109}, 10787 (2012)

\bibitem{pawar08}
A.B. Pawar, I.~Kretzschmar, Langmuir \textbf{24}, 355 (2008)

\bibitem{chen11b}
Q.~Chen, E.~Diesel, J.K. Whitmer, S.C. Bae, E.~Luijten, S.~Granick, J. Am.
  Chem. Soc. \textbf{133}, 7725 (2011)

\bibitem{sacanna10}
S.~Sacanna, W.T.M. Irvine, P.M. Chaikin, D.J. Pine, Nature \textbf{464}, 575
  (2010)

\bibitem{manoharan03}
V.N. Manoharan, M.T. Elsesser, D.J. Pine, Science \textbf{301}, 483 (2003)

\bibitem{jiang10}
S.~Jiang, Q.~Chen, M.~Tripathy, E.~Luijten, K.S. Schweizer, S.~Granick, Adv.
  Mater. \textbf{22}, 1060 (2010)

\bibitem{nelson02}
D.R. Nelson, Nano Letters \textbf{2}, 1125 (2002)

\bibitem{yethiraj03}
A.~Yethiraj, A.~van Blaaderen, Nature \textbf{421}, 513 (2003)

\bibitem{osterman09}
N.~Osterman, I.~Poberaj, J.~Dobnikar, D.~Frenkel, P.~Ziherl, D.~Babi{\'c},
  Phys. Rev. Lett. \textbf{103}, 228301 (2009)

\bibitem{smallenburg12}
F.~Smallenburg, H.R. Vutukuri, A.~Imhof, A.~van Blaaderen, M.~Dijkstra, J.
  Phys.: Condens. Matter \textbf{24}, 464113 (2012)

\bibitem{bliu13}
B.~Liu, T.H. Besseling, M.~Hermes, A.F. Demir{\"o}rs, A.~Imhof, A.~van
  Blaaderen, Nature Comm. \textbf{5}, 3092 (2014)

\bibitem{crassous14}
J.J. Crassous, A.M. Mihut, E.~Wernersson, P.~Pfleiderer, J.~Vermant, P.~Linse,
  P.~Schurtenberger, Nature Comm. \textbf{5}, 5516 (2014)

\bibitem{bharti15}
B.~Bharti, O.D. Velev, Langmuir \textbf{31}, 7897 (2015)

\bibitem{leunissen09}
M.E. Leunissen, H.R. Vutukuri, A.~van Blaaderen, Adv. Mater. \textbf{21}, 3116
  (2009)

\bibitem{martin03}
J.E. Martin, R.A. Anderson, R.L. Williamson, J. Chem. Phys. \textbf{118}, 1557
  (2003)

\bibitem{gangwal10}
S.~Gangwal, A.~Pawar, I.~Kretzschmar, O.D. Velev, Soft Matt. \textbf{6}, 1413
  (2010)

\bibitem{yan12}
J.~Yan, M.~Bloom, S.C. Bae, E.~Luijten, S.~Granick, Nature \textbf{491}, 578
  (2012)

\bibitem{yan15}
J.~Yan, S.C. Bae, S.~Granick, Adv. Mater. \textbf{27}, 874 (2015)

\bibitem{song15}
P.~Song, Y.~Wang, Y.~Wang, A.D. Hollingsworth, M.~Weck, D.J. Pine, M.D. Ward,
  J. Am. Chem. Soc. \textbf{137}, 3069 (2015)

\bibitem{levitt78}
D.G. Levitt, Biophys. J. \textbf{22}, 209 (1978)

\bibitem{zauhar85}
R.J. Zauhar, R.S. Morgan, J. Mol. Biol. \textbf{186}, 815 (1985)

\bibitem{allen01}
R.~Allen, J.P. Hansen, S.~Melchionna, Phys. Chem. Chem. Phys. \textbf{3}, 4177
  (2001)

\bibitem{boda04}
D.~Boda, D.~Gillespie, W.~Nonner, D.~Henderson, B.~Eisenberg, Phys. Rev. E
  \textbf{69}, 046702 (2004)

\bibitem{jadhao12}
V.~Jadhao, F.J. Solis, M.~Olvera de~la Cruz, Phys. Rev. Lett. \textbf{109},
  223905 (2012)

\bibitem{barros14a}
K.~Barros, D.~Sinkovits, E.~Luijten, J. Chem. Phys. \textbf{140}, 064903 (2014)

\bibitem{barros14b}
K.~Barros, E.~Luijten, Phys. Rev. Lett. \textbf{113}, 017801 (2014)

\bibitem{crc2015}
W.H. Haynes, ed., \emph{CRC Handbook of Chemistry and Physics}, 96th~edn. (CRC
  Press, Boca Baton, FL, 2015)

\bibitem{butcher91}
P.N. Butcher, D.~Cotter, \emph{The elements of nonlinear optics} (Cambridge
  University Press, Cambridge, U.K., 1990)

\bibitem{wu16a}
H.~Wu, M.~Han, E.~Luijten, \emph{Dielectric effects on the electric double
  layer of a {J}anus colloid} (2016)

\bibitem{long98}
D.~Long, A.~Ajdari, Phys. Rev. Lett. \textbf{81}, 1529 (1998)

\bibitem{levitan05}
J.A. Levitan, S.~Devasenathipathy, V.~Studer, Y.~Ben, T.~Thorsen, T.M. Squires,
  M.Z. Bazant, Colloids Surf. A \textbf{267}, 122 (2005)

\bibitem{bazant06}
M.Z. Bazant, Y.~Ben, Lab Chip \textbf{6}, 1455 (2006)

\bibitem{squires06}
T.M. Squires, M.Z. Bazant, J. Fluid Mech. \textbf{560}, 65 (2006)

\bibitem{bazant09}
M.Z. Bazant, M.S. Kilic, B.D. Storey, A.~Ajdari, Adv. Colloid Interface Sci.
  \textbf{152}, 48 (2009)

\bibitem{jackson99}
J.D. Jackson, \emph{Classical Electrodynamics}, 3rd~edn. (Wiley, New York,
  1999)

\bibitem{mangelsdorf97}
C.S. Mangelsdorf, L.R. White, Trans. Faraday Soc. \textbf{93}, 3145 (1997)

\bibitem{harned47}
H.S. Harned, R.L. Nuttall, J. Am. Chem. Soc. \textbf{69}, 736 (1947)

\bibitem{zerrouki08}
D.~Zerrouki, J.~Baudry, D.~Pine, P.~Chaikin, J.~Bibette, Nature \textbf{455},
  380 (2008)

\end{thebibliography}

\end{document}